\documentclass[preprint,12pt, a4paper]{elsarticle}

\usepackage{amssymb}
\usepackage{amsmath}
\usepackage{hyperref}
\usepackage{subfig}
\usepackage{booktabs}
\usepackage{makecell}

\usepackage{listings}
\usepackage{xcolor}
\definecolor{codegreen}{rgb}{0,0.6,0}
\definecolor{codegray}{rgb}{0.5,0.5,0.5}
\definecolor{codepurple}{rgb}{0.58,0,0.82}
\definecolor{backcolour}{rgb}{0.95,0.95,0.92}

\lstdefinestyle{mystyle}{
    backgroundcolor=\color{backcolour},   
    commentstyle=\color{codegreen},
    keywordstyle=\color{magenta},
    numberstyle=\tiny\color{codegray},
    stringstyle=\color{codepurple},
    basicstyle=\ttfamily\footnotesize,
    breakatwhitespace=false,         
    breaklines=true,                 
    captionpos=b,                    
    keepspaces=true,                 
    numbers=left,                    
    numbersep=5pt,                  
    showspaces=false,                
    showstringspaces=false,
    showtabs=false,                  
    tabsize=2
}
\journal{SoftwareX}

\begin{document}
\renewcommand{\labelenumii}{\arabic{enumi}.\arabic{enumii}}

\begin{frontmatter}

\title{dqc\_simulator: an easy-to-use distributed quantum computing simulator}


\author[label1]{Kenny Campbell}
\address[label1]{Heriot-Watt University, Edinburgh EH14 4AS, kenny.campbell@hw.ac.uk} 

\begin{abstract}
Distributed quantum computing (DQC) is a promising proposal for overcoming the scalability challenges of quantum computing. However, the evaluation of DQC hardware and software is difficult due to the relative dearth of classical simulation tools available for DQC devices. In this work, we introduce dqc\_simulator, a novel simulation toolkit, written in Python, which automates many of the most challenging aspects of the DQC simulation workflow. dqc\_simulator enables the easy simulation of both hardware and software, making it easy to create realistic and robust tests and benchmarks for full DQC systems.



\end{abstract}


\end{frontmatter}


\section*{Metadata}


\begin{table}[!h]
\begin{tabular}{|l|p{6.5cm}|p{6.5cm}|}
\hline
\textbf{Nr.} & \textbf{Code metadata description} & \textbf{Metadata} \\
\hline
C1 & Current code version & v0.2.6 \\
\hline
C2 & Permanent link to code/repository used for this code version & \url{https://github.com/km-campbell/dqc_simulator.git} \\
\hline
C3  & Permanent link to Reproducible Capsule & \\
\hline
C4 & Legal Code License   & Apache License, version 2.0 \\
\hline
C5 & Code versioning system used & git \\
\hline
C6 & Software code languages, tools, and services used & Python \\
\hline
C7 & Compilation requirements, operating environments \& dependencies & Only operates in Linux (tested in Ubuntu 20.04) and MacOS 64-bit (x84\_64). Dependencies: Python 3.9, NetSquid 1.1.7, NetSquid-PhysLayer 4.2.0, PyParsing 3.0.9  \\
\hline
C8 & If available Link to developer documentation/manual & \url{https://km-campbell-dqc-simulator.readthedocs.io/en/latest/} \\
\hline
C9 & Support email for questions & kennycampbell1@btinternet.com  \\
\hline
\end{tabular}
\caption{Code metadata}
\label{codeMetadata} 
\end{table}




\section{Motivation and significance}\label{sec:motivation_and_significance}

Quantum computing promises exponential speed-ups for some algorithms over conventional, classical, computers. However, a number of challenges must be overcome before an exponential advantage can be brought to bear on useful, real-world problems. In particular, the number of physical qubits must be increased to hundreds of thousands or millions of high-quality qubits \cite{webster_pinnacle_2026, gidney_how_2025}, instead of the hundreds of qubits present on existing devices.  Distributed quantum computing (DQC) is increasingly emerging as a way of bridging this scalability gap \cite{caleffi_distributed_2024}, but the field is less mature than monolithic, single-processor, quantum computing, and lacks some of the classical computing tools needed to accelerate research \cite{caleffi_distributed_2024}.

Quantum computers are extremely expensive to build and so there is a need for classical simulators to evaluate proposed architectures, compilers and tools for DQC. However, prior to this work, there were no publicly available noisy DQC simulators available that realistically account for the role of time, making it extremely time consuming to understand the role of realistic noise in DQC systems. There is some prior work exploring noiseless DQC simulation \cite{vazquez-perez_cunqa_2025, dqcExecutor}, but this makes it difficult to fully evaluate or benchmark realistic DQC hardware, in which noise will always be present. Furthermore, Ref. \cite{dqcExecutor} remains in the early stages of development. There are also two simulators, SimDisQ \cite{zhang_end--end_2025} and DQCS \cite{DQCS}, which do include some noise modelling but do not account for realistic time-dependent noise, which is necessary to fully categorise the noise in any quantum network. Furthermore, the  SimDisQ package does not yet seem to be publicly available and we were unable to find it on any public repository, such as GitHub, or package registry like PyPi. 

There are also several more general quantum network simulators \cite{coopmans_netsquid_2021, wu_sequence_2021, bartlett_distributed_2018, Satoh_QuISP_2021, diadamo_qunetsim_2020} intended for the design and evaluation of quantum internet applications. Although the time-dependent noisy simulation of distributed quantum computing devices is theoretically possible on some of these simulators, specifically NetSquid \cite{coopmans_netsquid_2021} and SeQUeNCe \cite{wu_sequence_2021}, realistic noisy simulation of benchmarking suites featuring many quantum circuits is unlikely to be practically feasible within a reasonable development timeline. To do so, each circuit would have to be manually split into many smaller subcomponents acting on different spatial network nodes and time steps. The quantum and classical messaging between nodes would also have to be manually specified at each occurrence, along with the management of networking resources such as communication qubits, the ancilla qubits used to facilitate quantum communication in DQC. Moreover, manual development of this kind would be extremely inflexible and make it challenging to make alterations to a given circuit being considered.

In this work, we introduce dqc\_simulator, an easy-to-use, open-source, discrete-event, DQC simulator that uses NetSquid under the hood to achieve powerful and flexible simulation of DQC hardware and software. With dqc\_simulator, users can easily specify quantum circuits and/or hardware for running on simulated DQC devices in the presence of realistic noise modelling. 

dqc\_simulator has already been used to perform novel evaluations in quantum data centres \cite{my_first_paper, my_second_paper}, a specific-type of distributed quantum computer, and aims to provide a tool that will be useful for the wider DQC and quantum networking research communities.



\section{Software description}


\subsection{Software architecture}

dqc\_simulator is written in Python 3.9 and makes heavy use of the NetSquid \cite{coopmans_netsquid_2021} library and much more limited use of the PyDynAA package, which underlies some of NetSquid's functionality. The parsing of .qasm files is done using modified code from the open-source nuqasm2 library and uses PyParsing. Finally, NumPy is used, as well as modules from the Python standard library. Due to restrictions imposed by the underlying NetSquid package, dqc\_simulator is only compatible with the Linux or MacOS 64-bit (x86\_64) operating systems, although virtualised Linux environments, such as Window's WSL2, can be used to circumvent this limitation.

\begin{figure}
    \centering
    \includegraphics[]{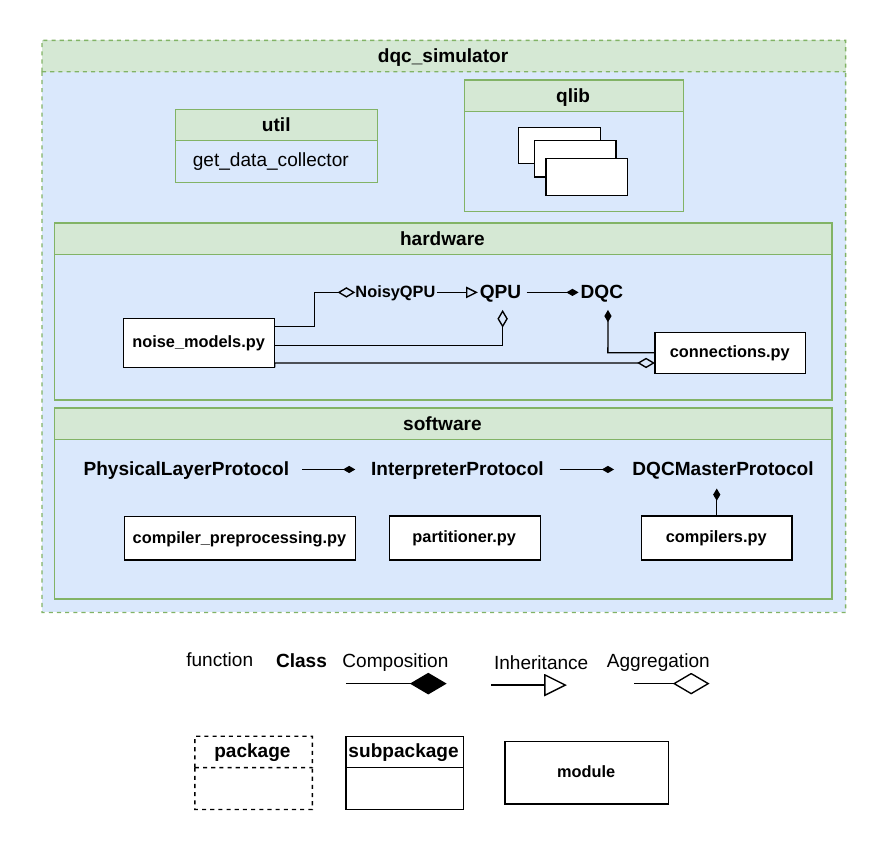}
    \caption{A simplified overview of the software architecture, showing the parts of the package relevant to the user. Many of the internal implementation details are omitted for clarity. Any arrows going from a module indicate the relationship between classes in the module to the object at the head of the arrow. This is done for brevity when there are multiple classes within the module that the user could use in a similar way. The arrows used have their standard Unified Modelling Language (UML) meanings \cite{uml2_spec}. Let class 1, $C_1$, be the class, at the tail of the arrow and class 2, $C_2$, be the class at the head of the arrow. Composition indicates that $C_2$ \textbf{has} a $C_1$. Inheritance means that $C_1$ \textbf{is} a $C_2$ and extends its functionality. Aggregation means that $C_2$ \textbf{can have} a $C_1$ but $C_1$ and $C_2$ can exist independently of each other.}
    \label{fig:software_architecture}
\end{figure}

A basic pictorial overview of the software architecture is shown in Fig. \ref{fig:software_architecture}.%
As shown in Fig. \ref{fig:software_architecture}, dqc\_simulator is organised into four subpackages: hardware, qlib, software, and util. The hardware subpackage provides classes representing DQC components such as QPUs and the connections between them, as well as an overall \lstinline{DQC} wrapper class that allows for the easy instantiation of an entire distributed quantum computer, given user provided details about the number and type of subcomponents to use and the topology of the network. The qlib subpackage is an extensible library of useful quantum circuits and identities; quantum gates; and states; along with some internally useful macros and information, such as various resource parameters for the remote gate schemes supported by the package. It is hoped that users will extend the constituent modules within qlib to include things that they have found useful during their own research. The software subpackage takes care of everything needed to go from a quantum circuit to something that the simulator can understand and simulate. It defines everything needed to either parse external .qasm files or define your own monolithic or distributed quantum circuit, partition a monolithic quantum circuit between different QPUs, and compile then interpret distributed quantum circuits. Finally, util provides helper functions that make development easier and aid the collection of simulation data.

 \subsection{Software functionalities}

The idea of dqc\_simulator is to retain as much as possible of the flexibility, generality and power of NetSquid while automating and simplifying much of the workflow to simulate distributed quantum computers. As such, dqc\_simulator can be used to simulate the hardware and/or software of distributed quantum computers by themselves or distributed quantum computers that exist as part of a larger and more general quantum networks, which can be simulated using NetSquid. The main features offered by dqc\_simulator are:
\begin{itemize}
    \item An easy-to-use interface for specifying distributed quantum circuits.
    \item The automatic partitioning and compilation of monolithic circuits between emulated quantum processing units (QPUs). 
    \begin{itemize}
        \item Monolithic circuits can be parsed from imported .qasm files \footnote{The parsing of .qasm files uses parsing functionality very heavily based on nuqasm2 \cite{nuqasm2} and uses a lightly modified version of modules from the nuqasm2 source code. This usage is consistent with the open-source license of nuqasm2.} or specified manually in Python.
        \item Users can also specify their own partitioning algorithms.
    \end{itemize}
    \item Automatic handling of communication qubits.
    \item An interpreter for commonly used communication subroutines such as remote gates. 
    \item Extensible libraries of pre-made compilers and compilation tools. 
        \begin{itemize}
            \item Users can also easily specify their own compilers, using the provided helper functions.
        \end{itemize}
    \item Full support for arbitrary noise, including time-dependent noise, using NetSquid's discrete-event simulation framework.
    \item Compatibility with all of the quantum formalisms offered by NetSquid. The formalism offered are ket states, density matrices (dense and sparse), the stabiliser formalism and graph states with local Cliffords \cite{coopmans_netsquid_2021}. 
        \begin{itemize}
            \item Users can specify the formalism used by changing one line of code, leaving all else intact.
        \end{itemize}
\end{itemize}
  

\section{Illustrative example: a case study}\label{sec:illustrative_examples}

To illustrate the power and simplicity of using dqc\_simulator, let's consider a specific example of a useful experiment which illustrates how to construct simulated DQC hardware; how to set up the quantum software; and how to run the simulation. Consider a researcher who wishes to predict how well different quantum algorithms would perform on DQC hardware that they may or may not have already created. This will help them to understand what possible experiments they might be able to run on their hardware and decide if the parameters that they can or expect to be able to deliver are sufficient. Simulation carries a much lower time cost than running the experiment and doing full state tomography over many shots only to find out that the noise is too high to converge to the correct answer. The researcher, referred to henceforth as the user, may wish to utilise an established external benchmarking suite to achieve this. All of this can be done relatively simply with dqc\_simulator by combining Listings \ref{lst:hardware}, \ref{lst:software} and \ref{lst:full_experiment} below. Some alternative options for specifying the quantum circuits are also included in Listing \ref{lst:alt_software}, although we consider a simpler experiment there, unrelated to the main case study explored in this section.

The code in Listings \ref{lst:hardware}, \ref{lst:software} and \ref{lst:full_experiment} emulates a three-QPU distributed quantum computer with user-specifiable noise parameters and then runs three different circuits on it. The code finds the fidelity of each circuit on the noisy device compared to the ideal, noiseless, implementation of the same circuit. The circuits considered are a GHZ generation circuit, Grover's algorithm and the Quantum Fourier Transform. Each circuit is specified as an external file written in openQASM 2.0. The circuit files are taken from the MQT Bench benchmarking suite \cite{quetschlich_mqt_2023}. The exact circuit files used are available in the examples directory within the dqc\_simulator GitHub repository along with the code in Listings \ref{lst:hardware}, \ref{lst:software} and \ref{lst:full_experiment}. 

For clarity of discussion, we split the experimental script into the aforementioned three listings here, however, the combined script is included in the examples directory of dqc\_simulator GitHub repository listed in Table \ref{codeMetadata} under the file name multiple\_monolithic\_circuits.py. Also included in the same directory are the aforementioned .qasm files needed to run the experiment. Finally, the code from Listings \ref{lst:hardware}, \ref{lst:software} and \ref{lst:full_experiment} is included in the same directory under the names setup\_hardware.py, setup\_sim.py, and run\_experiment.py. Cloning the repository, following the installation instructions and then running either multiple\_monolithic\_circuits.py or run\_experiment.py from within the examples directory should reproduce the expected results commented out at the end of Listing \ref{lst:full_experiment}.


To begin building the experiment, the user first sets up the hardware as shown in Listing \ref{lst:hardware}.

\begin{lstlisting}[language=Python, caption=Setting up simulated DQC hardware., label=lst:hardware]
import itertools as it

from dqc_simulator.hardware.connections import BlackBoxEntanglingQsourceConnection
from dqc_simulator.hardware.dqc_creation import DQC
from dqc_simulator.hardware.quantum_processors import NoisyQPU
from dqc_simulator.qlib.states import werner_state


def setup_hardware(
    F_werner=1,
    p_depolar_error_cnot=0,
    single_qubit_gate_error_prob=0,
    meas_error_prob=0,
    memory_depolar_rate=0,
):
    ent_dist_rate = 182  # Hz

    # Defining QPU
    qpu_class = NoisyQPU
    kwargs4qpu = {
        "p_depolar_error_cnot": p_depolar_error_cnot,
        "single_qubit_gate_error_prob": single_qubit_gate_error_prob,
        "meas_error_prob": meas_error_prob,
        "comm_qubit_depolar_rate": memory_depolar_rate,
        "proc_qubit_depolar_rate": memory_depolar_rate,
        "single_qubit_gate_time": 135 * 10**3,
        "two_qubit_gate_time": 600 * 10**3,
        "measurement_time": 600 * 10**4,
        "num_positions": 10,
        "num_comm_qubits": 2,
    }

    # Defining connection
    entangling_connection_class = BlackBoxEntanglingQsourceConnection
    kwargs4conn = {
        "delay": 1e9 / ent_dist_rate,  # 1e9 used because ent_dist_rate in Hz
        "state4distribution": werner_state(F_werner),
    }

    # Setting up the hardware
    num_qpus = 3
    quantum_topology = list(it.combinations(range(3), 2))
    classical_topology = list(it.combinations(range(3), 2))
    dqc = DQC(
        entangling_connection_class,
        num_qpus,
        quantum_topology,
        classical_topology,
        qpu_class=qpu_class,
        **kwargs4qpu,
        **kwargs4conn,
    )
    return dqc

\end{lstlisting}

Listing \ref{lst:hardware} illustrates the key steps for emulating realistic quantum hardware in dqc\_simulator. Users first define individual experimental components to emulate and then wrap them together with the \lstinline{DQC} class, which makes it easy to combine the emulated experimental components into a topology of the user's choice. Pre-made emulated experimental components are provided, with tunable parameters, but the user can also specify their own components by subclassing from the provided base classes. Some base classes are provided by dqc\_simulator and others by the underlying NetSquid package. For example, to define an emulated QPU, users can use the \lstinline{QPU} base class provided by dqc\_simulator, while to define a connection between QPUs, users can turn to the \lstinline{Connection} class provided by NetSquid. In the long term, it is hoped that users will add their own pre-made experimental components to dqc\_simulator. 

Users can then define a quantum algorithm to run on the emulated hardware that they have created, as shown in Listing \ref{lst:software}. There are two main ways of doing this. The first option, which is depicted in Listing \ref{lst:software}, is that the user starts from a monolithic circuit and use tools in the software.compiler\_preprocessing and software.partitioner modules to obtain a distributed quantum circuit. The second option is to manually specify a distributed quantum circuit directly, which may be more convenient for very small quantum circuits. 

For the first option, the user provides a monolithic quantum circuit by importing an openQASM 2.0 \cite{OpenQASM2.0_paper} file or defining a list of tuples with the form (gate type, qubit index, ...), where the ellipsis indicates that further qubit indices can be specified for multi-qubit gates. The workflow for working with external .qasm files is nicely exemplified by the \lstinline{setup_sim} function in Listing \ref{lst:software}. Firstly, we parse and translate the .qasm file into a form that dqc\_simulator can work with, using the \lstinline{preprocess_to_compilable_monolthic} function from the \lstinline{compiler_preprocessing} module. This produces a \lstinline{DqcCircuit} object, from which the required gate tuples can be extracted. From there, the qubits within the circuit are allocated to specific QPUs in the emulated hardware. In Listing \ref{lst:software}, this is done using the \lstinline{first_come_first_served_qubits_to_qpus} function from the software.partitioner module. After the allocation is decided, the previously defined monolithic circuit is partitioned into a distributed quantum circuit using \lstinline{partition_gate_tuples} from the software.partitioner module. The circuit is then in a form that can be readily simulated on the emulated hardware using \lstinline{DqcMasterProtocol} and will be compiled using the compiler function. In Listing \ref{lst:software}, we explicitly include the default option for systems with more than two QPUs, which is \lstinline{sort_many_qpus_greedily_by_node_and_time} from the software.compilers module. If no compiler is specified, this is what will be used unless the emulated hardware has only two QPUs, in which case \lstinline{sort_greedily_by_node_and_time} is used.

\begin{lstlisting}[language=Python, caption=Setting up the simulation to run distributed quantum circuits., label=lst:software]
from dqc_simulator.software.compilers import (
    sort_many_qpus_greedily_by_node_and_time as default_compiler,
)
from dqc_simulator.software.compiler_preprocessing import (
    preprocess_qasm_to_compilable_monolithic as preprocess,
)
from dqc_simulator.software.dqc_control import DQCMasterProtocol
from dqc_simulator.software.partitioner import (
    first_come_first_served_qubits_to_qpus as allocate,
    partition_gate_tuples as partition,
)


def setup_sim(dqc, circuit_filepath):
    # Retrieving QPU nodes from DQC
    nodes = list(dqc.nodes.values())

    # import .qasm file and convert to gate_tuples for monolithic_circuit
    include_path = "."  # assuming qelib1.inc is in current working directory
    monolithic_circuit = preprocess(circuit_filepath, include_path=include_path)
    monolithic_circuit = monolithic_circuit.ops  # gate_tuples

    # Determine allocation of processing qubits to QPUs
    old_to_new_lookup, proc_qubit_allocation4each_qpu = allocate(monolithic_circuit, list(dqc.nodes.values()))

    # Partition according to the previously defined qubit allocation
    scheme = "cat"  # the remote gate scheme to use
    partitioned_gate_tuples = partition(
        monolithic_circuit,
        dqc,
        scheme,
        old_to_new_lookup,
        proc_qubit_allocation4each_qpu,
    )
    # Setting up the software
    protocol = DQCMasterProtocol(
        partitioned_gate_tuples, nodes=dqc.nodes, 
        compiler_func=default_compiler
    )
    return protocol, nodes
\end{lstlisting}

A key facet of the workflow is that users can exchange their own policies for qubit allocation between QPUs and compilation. The capacity to do so is an integral part of the design of dqc\_simulator. Though such manual specification obviously requires more manual work than using the defaults, many users may be specifically trying to evaluate the performance of their novel partitioners or compilers and so this functionality is important to include. In each case, the API of the default functions serves as a guide for what is expected by the backend.

Alternatively, users can manually define a distributed quantum circuit by creating a list of gate tuples with the form (gate type, qubit index, node name, ..., remote gate type), where the ellipsis indicates that more qubit indices and node names can be added, in the same pattern, for multi-qubit gates. The node name refers to the name of a network node containing a QPU. The QPU in question must contain the qubit specified by the preceding qubit index. Network nodes are conventionally named node\_0, ..., node\_($n-1$), where $n$ is the number of nodes in the network. The allowed remote gate types are "cat", "1tp", "2tp" and "tp\_safe", which correspond to cat-comm, 1TP, 2TP and TP-safe from Ref. \cite{my_first_paper}. For local gates, a smaller gate tuple with no remote gate type entry is used. This alternative workflow is demonstrated for a single simple circuit in Listing \ref{lst:alt_software}. For simplicity and conciseness, this represents an unrelated experiment to the case study investigated in this section and is purely for illustrating this alternative workflow. It also illustrates a different method of data collection to the one used in the main case study.

\begin{lstlisting}[language=Python, caption=An alternative workflow for setting up the simulation to run distributed quantum circuits., label=lst:alt_software]
from netsquid.components import instructions as instr
import numpy as np

from dqc_simulator.software.dqc_control import DQCMasterProtocol
from dqc_simulator.util.helper import get_data_collector

def setup_sim(dqc):
    # Retrieving QPU nodes from DQC
    node_0 = dqc.get_node("node_0")
    node_1 = dqc.get_node("node_1")
    node_2 = dqc.get_node("node_2")

    # Identifying the processing qubits that we wish to initialise
    qubits0 = node_0.qmemory.processing_qubit_positions[0:3]
    qubits1 = node_1.qmemory.processing_qubit_positions[0:3]
    qubits2 = node_2.qmemory.processing_qubit_positions[0:3]

    # Defining the gates
    gate_tuples = [
        (instr.INSTR_INIT, qubits0, node_0.name),
        (instr.INSTR_INIT, qubits1, node_1.name),
        (instr.INSTR_INIT, qubits2, node_2.name),
        (instr.INSTR_H, qubits0[0], node_0.name),
        (instr.INSTR_CNOT, qubits0[0], node_0.name, qubits1[0], node_1.name, "cat"),
    ]

    # Setting up the software
    protocol = DQCMasterProtocol(gate_tuples, nodes=dqc.nodes)

    # Preparing data collection
    qubit_indices_2b_checked = [(qubits0[0], node_0), (qubits1[0], node_1)]
    desired_state = np.sqrt(1 / 2) * np.array([[1], [0], [0], [1]])
    dc = get_data_collector(protocol, qubit_indices_2b_checked, desired_state)
    return protocol, dc
\end{lstlisting}

However the distributed quantum circuit is arrived at, it is then fed into \lstinline{DQCMasterProtocol} along with some details from the hardware. \lstinline{DQCMasterProtocol} interprets the distributed quantum circuit. The quantum state after the simulation run is stored in the hardware objects within the simulation. Alternatively, a NetSquid \lstinline{DataCollector} can be used to preserve data from any point during the simulation. A simple premade \lstinline{DataCollector} for identifying the fidelity at the end of a quantum circuit relative to a specified state can be created using the get\_data\_collector function from the util/helper.py module, as shown in Listing \ref{lst:alt_software}. Alternatively, users running more complicated experiments can specify their own as discussed in the NetSquid documentation.

Once the emulated DQC hardware and software are both established, it is fairly simple to run the simulation, as shown in Listing \ref{lst:full_experiment}. It is worth noting that adjusting line 17 of Listing \ref{lst:full_experiment} will change the mathematical formalism used by the simulation. It is also worth noting that the final two import statements assume that Listings \ref{lst:hardware} and \ref{lst:software} were written in modules called setup\_hardware.py and setup\_sim.py, in the same directory as Listing \ref{lst:full_experiment}. One further subtlety of the code shown in Listing \ref{lst:full_experiment} is the use of the \lstinline{@processify} decorator imported from the util/helper.py module. \lstinline{processify} is a slightly modified reimplementation of the function available in the skywalker package \cite{skywalker}, and originally proposed in Ref. \cite{processify}. \lstinline{@processify} is intended to wrap functions, such as the \lstinline{get_fidelity} function, and forces the wrapped function to run inside its own subprocess. This is not necessary for the example shown here but is needed for multiple runs of larger experiments, due to a memory leak in the underlying NetSquid package dqc\_simulator relies on. The leak has been reported but, for now, the \lstinline{@processify} decorator provides a useful workaround for the issue.

\begin{lstlisting}[language=Python, caption=Running the case study experiment., label=lst:full_experiment]
import netsquid as ns
from netsquid.qubits import QFormalism, qubitapi as qapi
from dqc_simulator.util.helper import processify

from setup_hardware import setup_hardware
from setup_sim import setup_sim


def take_experimental_shot(
    circuit_filepath,
    F_werner=1,
    p_depolar_error_cnot=0,
    single_qubit_gate_error_prob=0,
    meas_error_prob=0,
    memory_depolar_rate=0,
):
    # Setting the formalism used to the density matrix formalism
    ns.set_qstate_formalism(QFormalism.DM)

    # Restting the state of the simulation (this is good practice)
    ns.sim_reset()

    # Setting up the hardware, software and data collection
    dqc = setup_hardware(
        F_werner=F_werner,
        p_depolar_error_cnot=p_depolar_error_cnot,
        single_qubit_gate_error_prob=single_qubit_gate_error_prob,
        meas_error_prob=meas_error_prob,
        memory_depolar_rate=memory_depolar_rate,
    )
    protocol, nodes = setup_sim(dqc, circuit_filepath)

    # Running the circuit
    protocol.start()
    ns.sim_run()

    # Identifying all initialised processing qubits
    qubits_2b_checked = []
    for node in nodes:
        positions = node.qmemory.processing_qubit_positions
        qubits = node.qmemory.pop(positions)
        qubits_2b_checked += [qubit for qubit in qubits if qubit is not None]
    return qubits_2b_checked

@processify
def get_fidelity(
                 circuit, 
                 F_werner,
                 p_depolar_error_cnot,
                 single_qubit_gate_error_prob,
                 meas_error_prob,
                 memory_depolar_rate):
    # Run ideal shot
    ideal_qubits = take_experimental_shot(circuit)
    actual_qubits = take_experimental_shot(
        circuit,
        F_werner=F_werner,
        p_depolar_error_cnot=p_depolar_error_cnot,
        single_qubit_gate_error_prob=single_qubit_gate_error_prob,
        meas_error_prob=meas_error_prob,
        memory_depolar_rate=memory_depolar_rate,
    )
    desired_state = qapi.reduced_dm(ideal_qubits)
    fidelity = qapi.fidelity(actual_qubits, desired_state, squared=True)
    return fidelity


def run_experiment(
    F_werner=1,
    p_depolar_error_cnot=0,
    single_qubit_gate_error_prob=0,
    meas_error_prob=0,
    memory_depolar_rate=0,
):
    # Choosing circuits to use (assuming the files are in the current working
    # directory)
    circuit_filepaths = [
        "ghz_5qubits.qasm",  # GHZ generation circuit
        "grover_5qubits.qasm",  # Grover algorithm
        "qft_5qubits.qasm",  # QFT
    ]

    data = {}
    for circuit in circuit_filepaths:
        fidelity = get_fidelity(
            circuit,
            F_werner=F_werner,
            p_depolar_error_cnot=p_depolar_error_cnot,
            single_qubit_gate_error_prob=single_qubit_gate_error_prob,
            meas_error_prob=meas_error_prob,
            memory_depolar_rate=memory_depolar_rate
        )
        data[circuit] = fidelity
    return data


print(
    run_experiment(
        F_werner=0.99,
        p_depolar_error_cnot=1e-03,
        single_qubit_gate_error_prob=2e-05,
        meas_error_prob=3e-03,
        memory_depolar_rate=0.055,
    )
)

# Expected result:
# {
# 'ghz_5qubits.qasm': 0.9570859020638329, 
# 'grover_5qubits.qasm': 0.10716586843690633, 
# 'qft_5qubits.qasm': 0.6470448636536242}
\end{lstlisting}


It is also worth noting that we have chosen a simple experimental experimental example for conciseness and to keep the runtime low, so that the results can quickly be verified by users. However, Listings \ref{lst:hardware}, \ref{lst:software} and \ref{lst:full_experiment} are easy to extend to other experiments. Including additional quantum circuits in the experiment is a trivial matter of adding the relevant file paths to the \lstinline{circuit_filepaths} list in the \lstinline{run_experiment} function. The addition of more nested for loops to the \lstinline{run_experiment} function in Listing \ref{lst:full_experiment} would allow users to vary over different noise parameters to gain an understanding of the impact that different types of noise would have on different quantum circuits or benchmark different quantum hardware. Both of these extensions are used for the experiments in Ref. \cite{my_first_paper} and the functionality of dqc\_simulator is central to taking the majority of the experimental results presented there.

Ref. \cite{my_first_paper} also verifies the correctness of dqc\_simulator by considering remote gates and comparing the results to analytical ones. Further validation of the correctness is carried out in the getting\_started script within the examples directory of dqc\_simulator's repository. The script can be used to check that we obtain the ideal GHZ state from a GHZ state generation circuit in the absence of noise. 

\section{Performance profiling}

To quantify the performance of dqc\_simulator, we profile the \lstinline{get_fidelity} function from Listing \ref{lst:full_experiment}, which carries out a single experimental shot for a single quantum circuit considered in the main example discussed in Sec. \ref{sec:illustrative_examples}. To investigate scalability, we extend the example to include different numbers of qubits for the same algorithms considered in Sec. \ref{sec:illustrative_examples}, namely GHZ state generation, Grover's algorithm, and the Quantum Fourier Transform (QFT). Two quantum formalisms are considered, the density matrix (DM) formalism and the stabiliser (STAB) formalism. For the latter, only the GHZ generation algorithm is considered because the stabiliser formalism cannot express non-Clifford circuits and the results are taken with lines 63 and 64 of Listing \ref{lst:full_experiment}, which compute the fidelity,  commented out. Instead, the fidelity is assigned to a NaN value because fidelity is ill defined in the stabiliser formalism and the fidelity result is not relevant to the profiling considered. The memory usage is profiled using the memory-profiler Python package \cite{memory_profiler} and the single shot runtime is profiled using the timeit package from the Python standard library. The full code used for profiling can be found in Ref. \cite{profiling_code} and the raw data corresponding to the results shown in this section can be found in Ref.  \cite{profiling_data}.  All profiling is done on a high-end desktop computer with 32Gb DDR5 RAM and an AMD Ryzen 5 7600X CPU.

\begin{figure}
    \centering
    \subfloat[]{\includegraphics[width=0.49\textwidth]{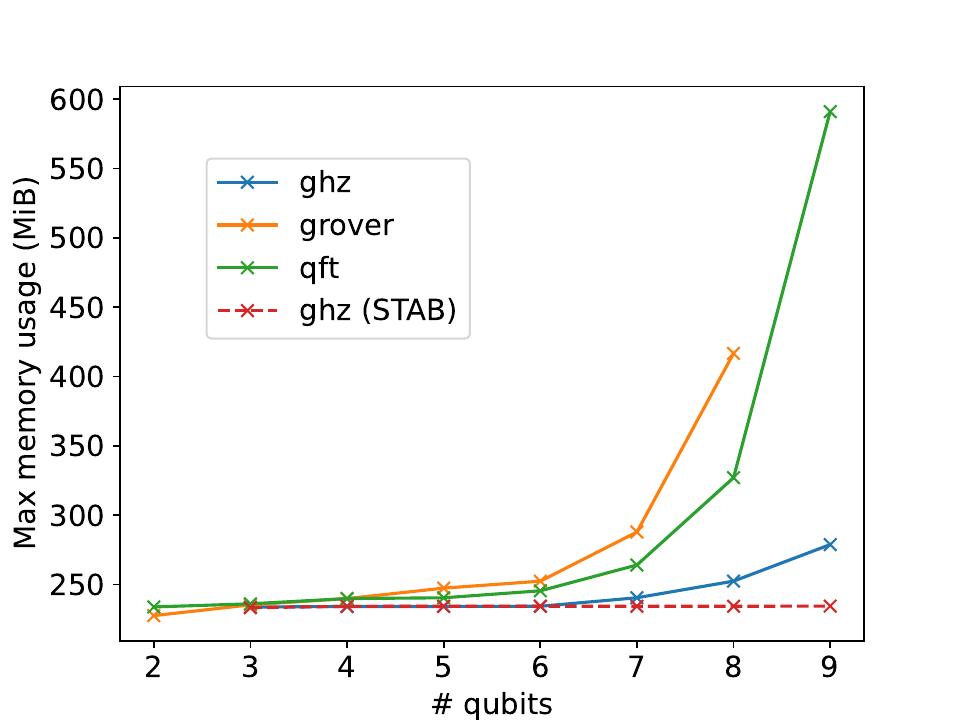}}
    \subfloat[]{\includegraphics[width=0.49\textwidth]{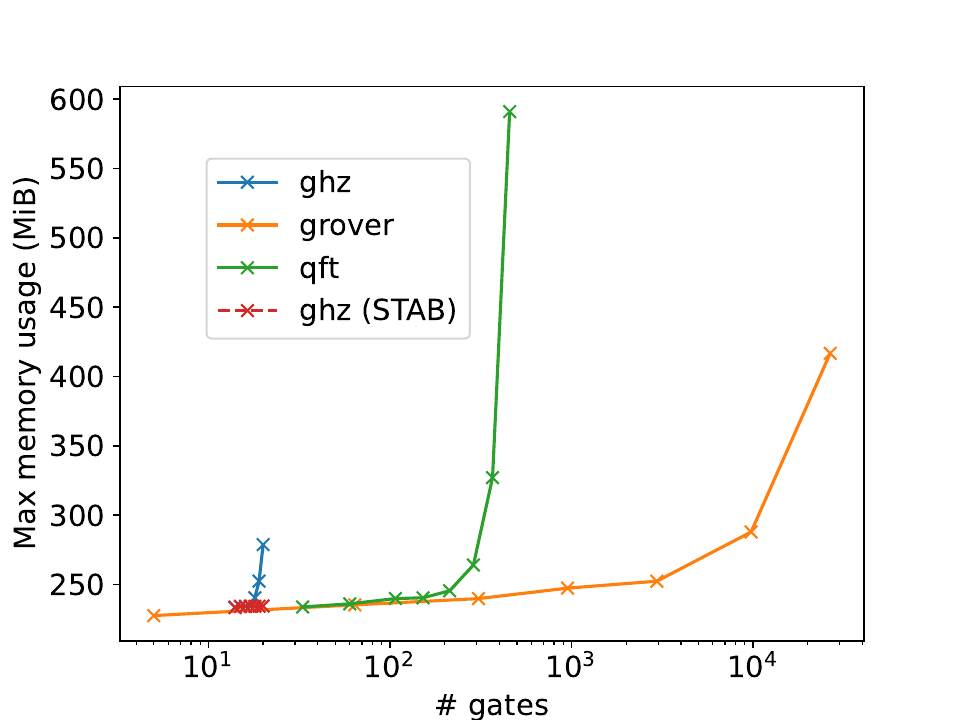}}
    \caption{The memory usage of different quantum circuits as a function of (a) the number of qubits, (b) the number of gates. In the legend, `ghz' refers to a GHZ generation algorithm, `grover' refers to Grover's algorithm, and `qft' refers to the Quantum Fourier Transform. `STAB` refers to results taken using the stabiliser formalism and all other results are taken in the density matrix formalism.}
    \label{fig:memory_usage}
\end{figure}

The maximimum memory usage obtained from one shot of various quantum circuits are plotted as a function of the number of qubits and the number of gates in Figs. \ref{fig:memory_usage}(a) and \ref{fig:memory_usage}(b), respectively.
From Fig. \ref{fig:memory_usage}, we can see that, as is typical, classical simulations using the density matrix formalism will only be useful for relatively small quantum circuits, due to the unfavourable exponential scaling of density matrix size with the number of qubits in the system. Even with just nine qubits and fewer than one thousand gates, we see that memory usage can be as high as 600 MiB and the availability of RAM will eventually limit the feasibility of simulations. That said, the much more efficiently scaling stabiliser formalism uses radically less memory as the number of qubits is increased, although, unlike the density matrix formalism, multiple shots would be needed to converge to a fidelity result.

\begin{figure}
    \centering
    \subfloat[]{\includegraphics[width=0.49\textwidth]{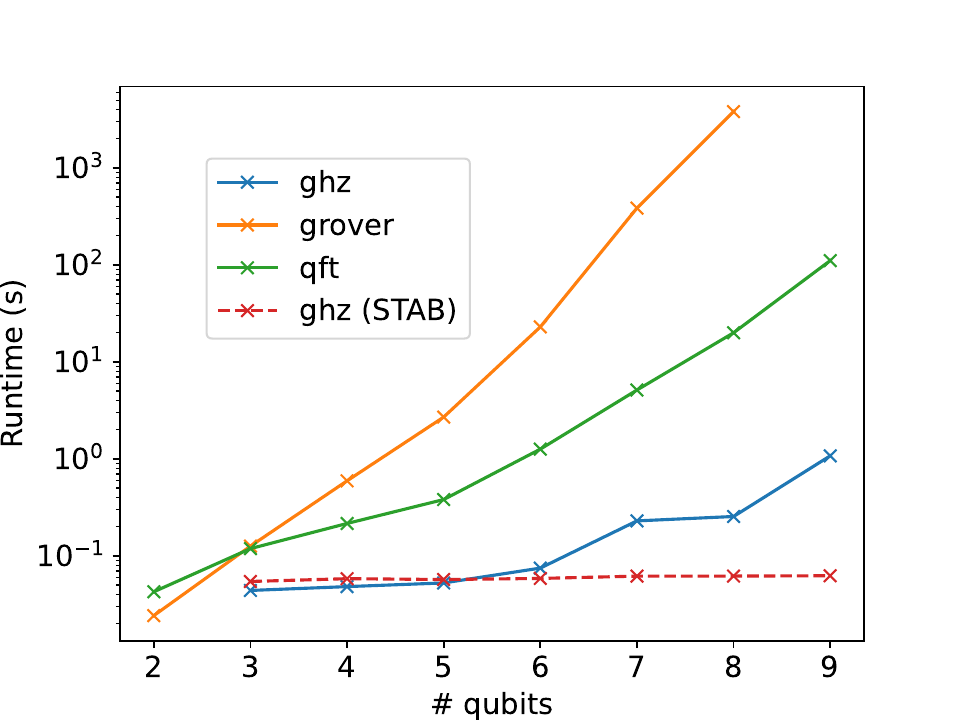}}
    \subfloat[]{\includegraphics[width=0.49\textwidth]{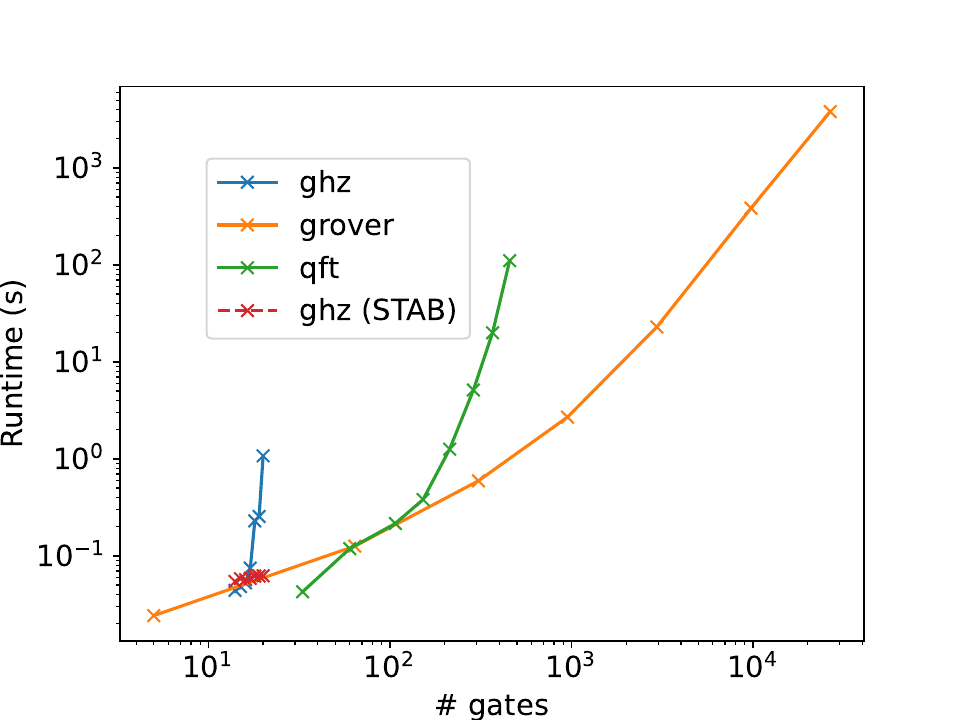}}
    \caption{The runtime for a single shot of different quantum circuits as a function of (a) the number of qubits, (b) the number of gates. In the legend, `ghz' refers to a GHZ generation algorithm, `grover' refers to Grover's algorithm, and `qft' refers to the Quantum Fourier Transform. `STAB' refers to results taken using the stabiliser formalism and all other results are taken in the density matrix formalism. Each data point is averaged over 50 runs other than data points for Grover's algorithm circuits with more than 4 qubits, for which 1 run is used.}
    \label{fig:runtime}
\end{figure}

The single-shot runtime results are shown as a function of the number of qubits and the number of gates in Figs. \ref{fig:runtime}(a) and \ref{fig:runtime}(b), respectively. Each data point is averaged over 50 runs except for implementations of Grover's algorithm with more than four qubits, for which one run is used. This is due to the large runtime of such circuits, which made using many iterations impractical and also was large enough that fluctuations in runtime are likely to be negligible.

The runtime values tell a similar story to the memory results. Again, runtime scales very rapidly with the number of qubits and gates in the circuit for the density matrix formalism and would quickly become a practical limitation on the circuit size considered. The stabiliser formalism allows for much more favourable scaling and makes it possible to consider much larger circuits but imposes the restriction that only Clifford gates can be considered.

\section{Impact}

dqc\_simulator is a broadly applicable tool for distributed quantum computing (DQC) research with applications in the evaluation of DQC hardware, architecture, compilation, and scheduling. The package can also be integrated into broader quantum network simulations that involve distributed quantum computers within them, aiding investigation into a quantum internet. Moreover, as mentioned in Sec. \ref{sec:illustrative_examples}, dqc\_simulator has already been used to conduct a detailed analysis on the impact of noise in DQCs \cite{my_first_paper}. Furthermore, in Ref. \cite{my_second_paper}, dqc\_simulator is used in conjuction with NetSquid to evaluate ways of combatting noise in the DQC setting \cite{my_second_paper}. 

We compare dqc\_simulator to existing tools in Table \ref{tab:comparision_with_other_sims}. Four dimensions are considered: the ability to carry out noise modelling; whether the simulator is a discrete-event simulator \cite{coopmans_netsquid_2021}, and thus able to precisely characterise timings; whether hardware is explicitly modelled; and whether circuit distribution is automatic, meaning that an inputted monolithic quantum circuit will be compiled into a single distributed quantum circuit. dqc\_simulator is the only simulator that has all of these features.

Overall, dqc\_simulator is distinguished from previous work by its ability to do detailed hardware modelling, including time-dependent noise, and still allow users to characterise algorithms with a single quantum circuit. As discussed in Ref. \cite{Dahlberg_simulaqron_2018}, precisely characterising timings and accounting for time-dependent noise is crucial for the accurate simulation of realistic quantum networks. Therefore, techniques like discrete-event simulation \cite{coopmans_netsquid_2021} or other ways of precisely accounting for the role of time are needed robustly model quantum networks. On the other hand, if users are unable to specify algorithms as a single quantum circuit, and must instead manually specify what every part of the network is doing at different times, then it is not feasible to benchmark emulated DQC hardware within a feasible time frame and small changes to the specification of an algorithm are difficult to make.

\begingroup 
\setlength{\tabcolsep}{10pt} 
\renewcommand{\arraystretch}{1} 
\begin{table}
    \centering
    \begin{tabular}{ccccc}\toprule
       Simulator & \thead{Noise \\ modelling} & \thead{Discrete\\ -event} & \thead{Hardware \\ modelling} & \thead{Circuit \\ distribution \\ automatic}  \\ \midrule
       NetSquid \cite{coopmans_netsquid_2021} & Yes & Yes & Yes & No\\ 
       SeQUeNCe \cite{wu_sequence_2021} & Yes & Yes & Yes & No  \\
       SQUANCH \cite{bartlett_distributed_2018} & Yes & No & No & No \\
       SimulaQron \cite{Dahlberg_simulaqron_2018} & QPU only & No & No & No \\
       QuISP \cite{Satoh_QuISP_2021} & Yes & Yes & No & No  \\
       QuNetSim \cite{diadamo_qunetsim_2020} & QPU only & No & No & No \\
       Interlin-q \cite{parekh_quantum_2021} & QPU only & No & No & Yes \\
       CUNQA \cite{vazquez-perez_cunqa_2025} & QPU only & No & No & No* \\
       DQCS \cite{DQCS} & Yes & No & No & Yes   \\
       dqc-executor \cite{dqcExecutor} & No & Yes & No & Yes  \\
       SimDisQ \cite{zhang_end--end_2025} & Limited* & No & No & Yes  \\
       This work & Yes & Yes & Yes & Yes \\
    \bottomrule
    \end{tabular}
    \caption{Comparison of different quantum network and DQC simulators. Here, the term `QPU only' in the noise modelling column indicates that noise is modelled only in the quantum processing units (QPUs) via third-party backends, meaning that there is no explicit modelling of any network noise. Such modelling could potentially be added in manually but is non-native. The asterix related to whether circuit distribution is automatic for CUNQA is to indicate that although one cannot start with a monolithic circuit and automatically distribute it, users can easily specify one distributed quantum circuit, rather than having to specify the low-level details of every interaction with the network. The asterix in the noise modelling entry for SimDisQ is to indicate that SimDisQ accounts for photon loss but no other link noise. We note that it is especially challenging to categorise dqc-executor \cite{dqcExecutor} because it is early in development and so the results presented for dqc-executor here are our best estimates for its functionality, based on the examples the developers provide and inspection of the source code.}
    \label{tab:comparision_with_other_sims}
\end{table}%
\endgroup%


\section{Conclusions}

In this work, we introduce dqc\_simulator, an easy-to-use simulation toolkit for distributed quantum computers. dqc\_simulator lowers the technical barrier to entry of simulating distributed quantum computers by automating the complicated details of hardware setup and software implementation, without preventing power users from specifying more complicated hardware and software if needed. 

The package is open source and includes extensive documentation, making it accessible to the entire research community and empowering researchers with the ability to easily simulate the systems that they want to study.

\section*{Acknowledgements}

I would like to thank Mohsen Razavi and Ahmed Lawey for allowing the freedom to pursue the development of this simulator and many invaluable discussions on quantum physics and software. I would like to thank Erika Andersson for useful discussion and comments. Finally, I would also like to acknowledge funding from UKRI grant UKRI598 and the Leeds Doctoral Scholarship.




\bibliographystyle{ieeetr} 
\raggedright
\bibliography{references.bib}








\end{document}